\documentclass{aastex61}

\newcommand\aastex{AAS\TeX}

\received{}
\revised{}
\accepted{}

\submitjournal{}

\shorttitle{\aastex\ Understanding Anisotropy with single source}
\shortauthors{X. B. Qu}

\begin{document}

\title{Understanding the galactic cosmic ray dipole anisotropy with a nearby single source under the spatially-dependent propagation scenario}

\correspondingauthor{Xiao-BO Qu}
\email{quxb@upc.edu.cn}

\author{Xiao-BO Qu}
\affiliation{College of Science, China University of Petroleum, Qingdao,  266555, China}

\begin{abstract}

Recently studies of the dipole anisotropy in the arrival directions of Galactic cosmic rays indicate that the TeV-PeV dipole anisotropy amplitude is not described by a simple power law, moreover a rapid phase change exists at an energy of $0.1\sim0.3$ PeV. In this work we argue that the dipole anisotropy amplitude and phase evolution with energies can be reproduced under the spatially-dependent propagation  scenario with a nearby single source added. Our results indicate a nearby single source have significant influence to the cosmic ray gradient below $0.1\sim0.3$ PeV under the spatially-dependent propagation scenario, which leads the dipole anisotropy phase change at this energy region. The dipole anisotropy amplitude of the galactic cosmic rays can also be maintained at a lower level, which are consistent with observations by underground muons and air shower experiments.

\end{abstract}


\section{Introduction} \label{sec:intro}

The anisotropy of cosmic rays (CRs) may be a precious tool to probe the propagation of CRs throughout the Galaxy \citep{bla12}. The CRs below $10^{17}$ eV are expected to be mainly Galactic, after released by their sources--presumably supernova remnants, they start to diffuse through the Galactic environment. As CRs are mostly charged nuclei, their paths are deflected and highly isotropized by the action of galactic magnetic field they propagate through before reaching the Earth atmosphere. Although the arrive direction is mainly isotropy, Galactic CRs possess a small but significant dipole anisotropy (DA) of order $10^{-4} \sim 10^{-3}$.  The origin of this large scale DA is still uncertain, but the study of its amplitude and phase evolution with energies has an important valence to understand the CRs propagation mechanisms.

The conventional propagation mechanism of CRs always derive a DA with amplitude much higher than observed. The conventional propagation models simply assume an isotropic diffusion with a smooth distribution of sources. In this case, the DA amplitude is expected to follow the energy scaling of the diffusion tensor. The DA amplitude derived by these models is always higher by about one order than the observation. Some models were proposed to reduce the amplitude of DA to the observed level, such as non-uniform distribution of sources\citep{bla12}, Spatially-dependent propagation(SDP)\citep{guo16}\citep{tom15}, Single nearby source plus conventional diffusion\citep{liu17}. In these models, SDP model can decrease the amplitude of DA by nearly an order of magnitude than that of the conventional propagation model, mainly by introduce an anti-correlation between the diffusion properties and the source distribution of CRs.

Despite that the amplitude of DA can be partially settled, the phase evolution with energies is difficult. The data from recent studies(ARGO-YBJ \citep{bar15}, EAS-TOP\citep{agl09}, IceCube/IceTop \citep{aar13}\citep{aar16}, and Tibet-AS \citep{ame17})indicate that the DA undergoes a rapid phase flip  (with almost vanishing amplitude)  at an energy of $0.1\sim0.3$ PeV, which means the CR density gradient has changed its direction at this energy. The phase evolution of the CR anisotropy with energy can not be explained by any large scale diffusion model without considering e.g., the local source and/or magnetic field effect\citep{ahl16}, because these diffusion models always give a DA with the direction of the Galactic center.

In this work, in order to explain the phase evolution of DA and simultaneous maintain the low amplitude according to the observation, we introduce a nearby single source under the scenario of SDP model. The SDP scenario leads to a low amplitude of DA, which means a small CR density gradient. So the single nearby source at nearly anti-galactic center direction can have significant influence to the CR density gradient, which means the DA phase will change its direction at this energy region. This letter is organized as follow: in Section 2, we introduce our model: including the main frame of the SDP scenario and the parameter sets of the single nearby source. In Section 3, we estimate the contribution of the single source to the CR density gradient and give the result of the DA amplitude and phase evolution with energies. In Section 4, we present the conclusion of this work.

\section{Model Description} \label{sec:model}

It is generally accepted that the CRs are accelerated at supernova remnants and then diffuse in the Galaxy. Before reaching the solar vicinity, the accelerated CRs may suffering from fragmentation and energy losses in the interstellar medium (ISM) and interstellar radiation field and magnetic field, decay and reacceleration or convection. The transport process of CRs in the can be described by the well-known diffusion equation:
\begin{equation}
\frac{\partial \psi (\vec{r},p,t)}{\partial t}  = Q(\vec{r},p,t)+\nabla \cdot (D_{xx} \nabla \psi )
 + \frac{\partial}{\partial p} p^2 D_{pp} \frac{\partial}{\partial p} \frac{1}{ p^2} \psi - \frac{\partial}{\partial p} [\dot{p} - \frac{p}{3} (\nabla \cdot V_{c} \psi  )]  -  \frac{\psi}{\tau_{f}} -  \frac{\psi}{\tau_{r}}
\label{pro}
\end{equation}
where $\psi (\vec{r},p,t)$ is the density of CR particles per unit momentum $p$ at position $\vec{r}$, $Q(\vec{r},p,t)$ is the source distribution, $D_{xx}$ is the spatial diffusion coefficient, $V_{c}$ is the convection velocity, $D_{pp}$ is the diffusive reacceleration coefficient in momentum space, $\dot{p}\equiv \mathrm{d}p / \mathrm{d} t$ is momentum loss rate, $\tau_{f}$ and $\tau_{r}$ are the characteristic time scales for fragmentation and radioactive decay respectively. $D_{pp}$ is also used to describe the reacceleration process, which is coupled with the spatial diffusion coefficient $D_{xx}$ as \citep{seo94}
\begin{equation}
\label{dpp}
D_{pp}D_{xx}=\frac{4p^{2}v_{A}^{2}}{3 \delta(4- \delta ^{2})(4- \delta) \omega}
\end{equation}
where $v_{A}$ is the Alfven speed, and $\omega$ is the ratio of magnetohydrodynamic wave energy density to the magnetic field energy density, which can be fixed to 1. The CRs propagate in an extended halo with a characteristic height $z_{h}$, beyond which free escape of CRs is assumed.

In this work, a nearby single source is added to the SDP scenario, and the others SNRs as the CR sources in the SDP model are named background SNRs.  For the background SNRs, which is already well introduced in the SDP model, it is viable to assume that the spatial distribution of CRs from them arrives at steady state. Nevertheless for the nearby single SNR, the time-dependent transport of CRs after injection is requisite\citep{liu17}. In the next subsections, we discuss these two components separately.

\subsection{Background supernova remnants}
The CRs density is only ascribed to the spatial diffusion coefficient $D_{xx}$ in SDP model, because $D_{pp}$ can be derived from $D_{xx}$ by Eq.~(\ref{dpp}) and the convection is ignored. The spatial diffusion coefficient in the SDP model is described with a two halo approach: the inner (disk) and outer halo. The half thickness of the halo is  $z_{h}$, the half thickness of inner halo is $\xi z_{h}$ and $(1- \xi) z_{h}$ is the half thickness of outer halo, For the inner halo, the diffusion coefficient is anti-correlated with the source distribution by means of a scale formula $F(r,z)$ fitted from the observed spatial distribution of SNRs. (details see \citet{guo16})
\begin{equation}
\label{D0}
D_{xx}(r,z,p)=F(r,z)D_{0}\beta \left(\frac{p}{p_{0}}\right)^{F(r,z)\delta_{0}}
\end{equation}
For the outer halo where the source term vanishes, the diffusion coefficient recovers the traditional form of $D_{0}\beta(p/p_{0})^{\delta_{0}}$  , where $p$ is rigidity and $D_{0}$ specifies its normalization at the reference rigidity $p_{0}$ and $\delta_{0}$ reflects the property of the ISM turbulence.

 A numerical method is necessary to solve the diffusion equations, especially in case that the diffusion varies everywhere in the Milky Way. In this work, we use the released DRAGON code\citep{evo08} to solve the CR propagation equation described in Eq.~(\ref{pro}). DRAGON allows us to perform a numerical calculation with a space-dependent coefficient mentioned above. The basic model parameters are given in Table~\ref{tab1}.

\begin{deluxetable}{cccccccccc}[b!]
\tablecaption{Parameters of the transport effects and background CR injection.\label{tab1}}
\tablecolumns{2}
\tablenum{1}
\tablewidth{0pt}
\tablehead{
\colhead{$D_{0} \ (\text{cm}^{2}\text{s}^{-1})$} &
\colhead{$\delta_{0}$}  &
\colhead{$v_{A}(\text{km s}^{-1})$}  &
\colhead{$z_{h}(\text{kpc})$}  &
\colhead{$\xi$}  &
\colhead{$log(q_{0})\tablenotemark{*}$}  &
\colhead{$\nu_{1}$}  &
\colhead{$\nu_{2}$}  &
\colhead{$p_{br}(\text{GV})$}  &
\colhead{$\hat{p}(\text{PV})$}
}
\startdata
$8.3 \times 10^{28}$  & 0.52 & 16  & 5 & 0.13 & -1.332 &  1.9 & 2.38 & 9.0 & 0.14 \\
\enddata
\tablenotetext{*}{Normalization at 100 GeV in units of cm$^{-2}$ s$^{-1}$ sr$^{-1}$ GeV$^{-1}$.}
\end{deluxetable}

 The injection spectrum of primary CRs at sources region for the background SNRs is taken as a broken power-law form:
\begin{equation}
\label{inj}
q(p)=q_{0} \times \Biggl\{
\begin{array}{cc}
(p/p_{br})^{-\nu_{1}}   & \ \text{if} (p<p_{br}) \\
(p/p_{br})^{-\nu_{2}} \cdot e^{-p/\hat{p}} &  \ \text{if} (p\geq p_{br}) ,
\end{array}
\end{equation}

where $p$ is the rigidity, $q_{0}$ is the normalization factor for all nuclei, and the relative abundance of each nuclei follows the default value in the DRAGON package. $p_{br}$ is the broken energy and $\nu_{1}$, $\nu_{2}$  are spectral index before and after the broken energy $p_{br}$. In this work the cutoff energy $\hat{p}$ is set to $1.4 \times 10^{5}$ GeV to fit the AMS-02+CREAM data\citep{guo18}. Detailed information of the parameters are listed in Table~\ref{tab1}.

\subsection{Nearby supernova remnant}

As mentioned above, diffusion model can't explain the DA amplitude and the phase evolution with energy. In the case of diffusion with a smooth distribution of sources, the DA is expected to simply align with the direction of the Galactic center. One possible reason which can cause the DA phase changes is a local nearby source.

We assume a local (300 pc from us) supernova explosion occurred about $10^{5}$ years ago in the direction of $(l=145^{\circ},b=0^{\circ})$. The charged particles were continually accelerated nearby the shock front with the expansion of supernova ejecta. The accelerated spectrum is represented by a power law plus an exponential cutoff, i.e.,
\begin{equation}
\label{single}
Q_{s}=q_{0}^{s}\mathcal{R}^{-\alpha}e^{-E/E_{cut}}.
\end{equation}
where $\mathcal{R}$ is the rigidity, $q_{0}^{s}$ is the normalization factor for nearby single source. $\alpha^{s}$ is  spectral index, $E_{cut}^{s}$ is the cutoff energy. Detailed information of the parameters are listed in Table~\ref{tab2}. The time-dependent distribution of CR nuclei from a point source can be evaluated by means of Green function technique, the corresponding analytical solution can be found in \citet{ber12}. The diffusion coefficient is set according to the SDP model inner halo setting, in the position of solar system the parameter  $F(r,z)$ in Eq.~\ref{D0} is 0.25.

\begin{deluxetable}{ccc}[b!]
\tablecaption{Injection spectrum of single nearby source. \label{tab2}}
\tablecolumns{2}
\tablenum{2}
\tablewidth{0pt}
\tablehead{
\colhead{$q_{0}^{s}[\text{GeV}^{-1}]$} &
\colhead{$\alpha^{s}$ }  &
\colhead{$E_{cut}^{s} [\text{GeV}]$}
}
\startdata
$5.5 \times 10^{52} $ & 2.33 & $ 0.55 \times 10^{5}$  \\
\enddata

\end{deluxetable}

\section{Calculation Results}\label{sec:results}

 According to the recently observation, the DA of CR have complicated behaviors, such as the DA phase change at $\sim 10^{14}$ eV,  and the DA amplitude evolution with energies is no more a simple power law feature. These new observation results can't explained by simple diffusion models.  In this work we introduce a nearby single source under the main frame of the SDP model to understand the cosmic ray DA.  The DRAGON code is used to calculate the galactic cosmic ray propagation, in which the essential transport parameters are $D_{0}$, $\delta$; the source parameters of background CRs are $p_{br}$, $\nu_{1}$, $\nu_{2}$, $\hat{p}$. For the local nearby SNR source, there are some parameters describing the CR injection spectra, i.e., $\alpha^s$, $E_{cut}^s$. In addition, to fit the low energy
data, we have to take solar modulation into account, which is represented by the modulation potential $\phi=600$ MV \citep{gle68}.

\subsection{The Amplitude and Phase of the Dipole Anisotropy}
After the nearby single source was added, the anisotropy can be understood as a sum of two components: one is caused by the background sources and the other is from the nearby single source. the total anisotropy can be calculated by:
\begin{equation}
\label{aniso}
\Delta = \frac{\sum\bar{I_{i}}\Delta_{i}n_{i} \cdot n_{\text{max}}}{\sum\bar{I_{i}}},
\end{equation}

 $n_{i}$ is the direction of the source and and $\Delta_{i}$ denotes the anisotropy of that source, $\bar{I_{i}}$ is the CR average intensity from that source. In this work, the sources are separated to two classes: background sources as a whole and a nearby single source. The CR intensity distribution can be get, after projected onto the  right ascension coordinate, the one-dimensional (1D) profile of the anisotropy is derived. The 1D profile of the anisotropy is
fitted by the first-order harmonic function in form of:
 \begin{equation}
\label{harm}
R(\alpha) = 1+A_{1}cos(\alpha - \phi_{1}),
\end{equation}
where $R(\alpha)$ denotes the relative intensity of CRs at right
ascension  $\alpha$, $A_{1}$ is the amplitude of the first harmonics, and $\phi_{1}$ is the phase at which $R(\alpha)$ reaches its maximum. Consider the observation from one experiment can't cover all declination, in this work below 10 TeV the declination cut is $-5^{\circ} \sim 65^{\circ}$ according to the ARGO-YBJ experiment, and above 10 TeV the declination cut is $-20^{\circ} \sim 80^{\circ}$ according to AS$\gamma$ experiment.

\begin{figure}[ht!]
\plotone{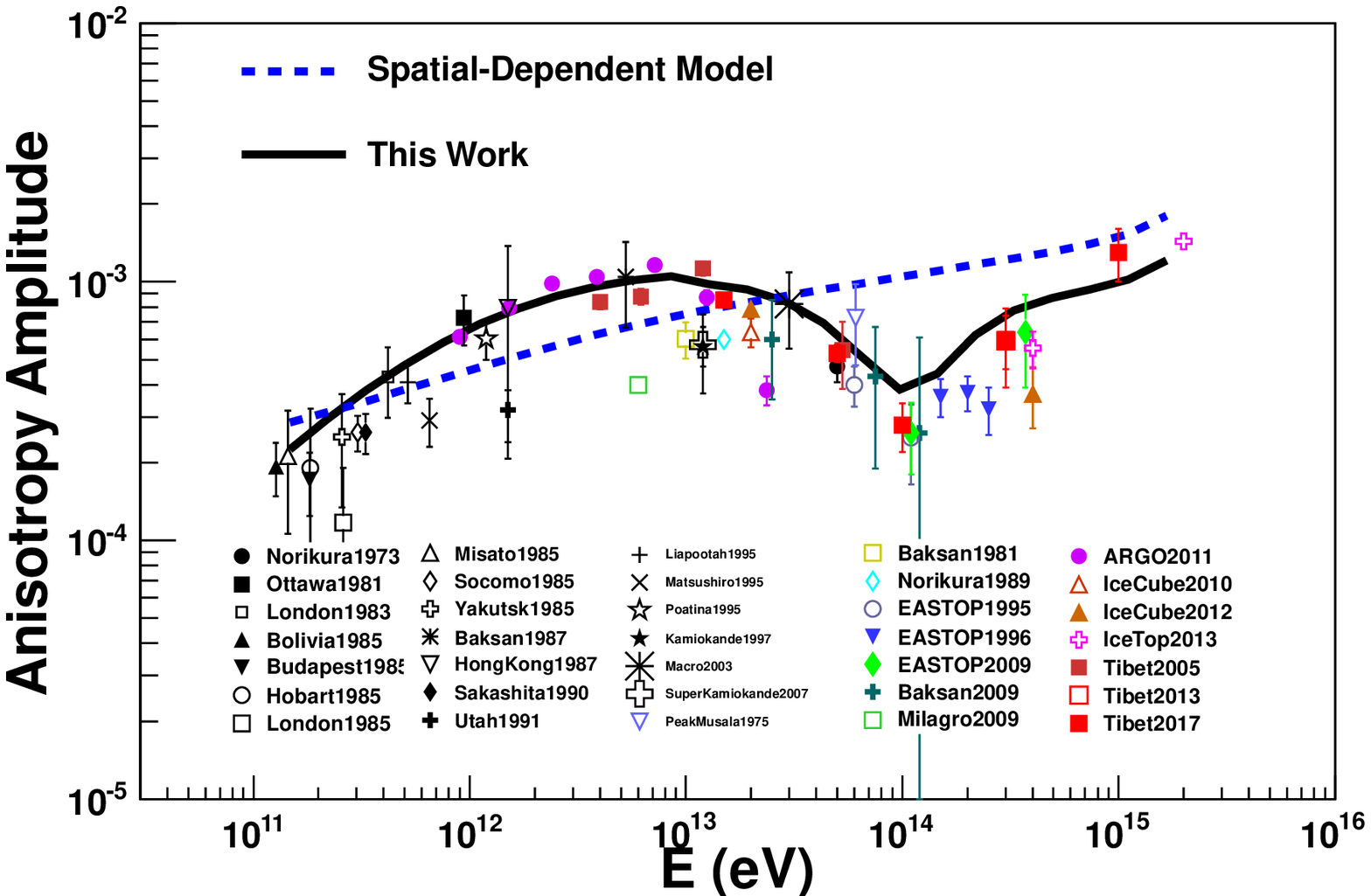}
\caption{\label{fig1} The amplitude of the DA expected from this work and the SPD model, compared
with the observations. The solid black line and blue dash line represent the model calculations in this work and the SDP model, respectively. The data come from underground muon observations from: Norikura (1973; \citep{sak73}), Ottawa
(1983; \citep{ber81}), London (1983; \citep{tha83}),
Bolivia (1985; \citep{swi85}), Budapest (1985; \citep{swi85}), Hobart (1985; \citep{swi85}), London (1985;
\citep{swi85}), Misato (1985; \citep{swi85}),
Socorro (1985; \citep{swi85}), Yakutsk (1985; \citep{swi85}), Banksan (1987; \citep{and87}), HongKong (1987;
\citep{lee87}), Sakashita (1990; \citep{uen90}), Utah (1991; \citep{cut91}), Liapootah (1995; \citep{mun95}), Matsushiro (1995;
\citep{mor95}), Poatina (1995; \citep{fen95}), Kamiokande (1997;
\citep{mun97}), Marco (2003; \citep{amb03}), SuperKamiokande
(2007; \citep{gui07}); and air shower array experiments from
PeakMusala (1975; \citep{gom75}), Baksan (1981; \citep{ale81}), EASTOP (1995, 1996, 2009; \citep{agl95}, \citep{agl96}, \citep{agl09}),
Baksan (2009; \citep{ale09}), Milagro(2009; \citep{abd09}),
ARGO (2011; \citep{iup11}), IceCube (2010, 2012; \citep{abb10}, \citep{abb12}),
IceTop (2013; \citep{aar13}), Tibet (2005, 2013, 2017; \citep{ame05}, \citep{ame13}, \citep{ame17} ).}
\end{figure}

\begin{figure}[ht!]
\plotone{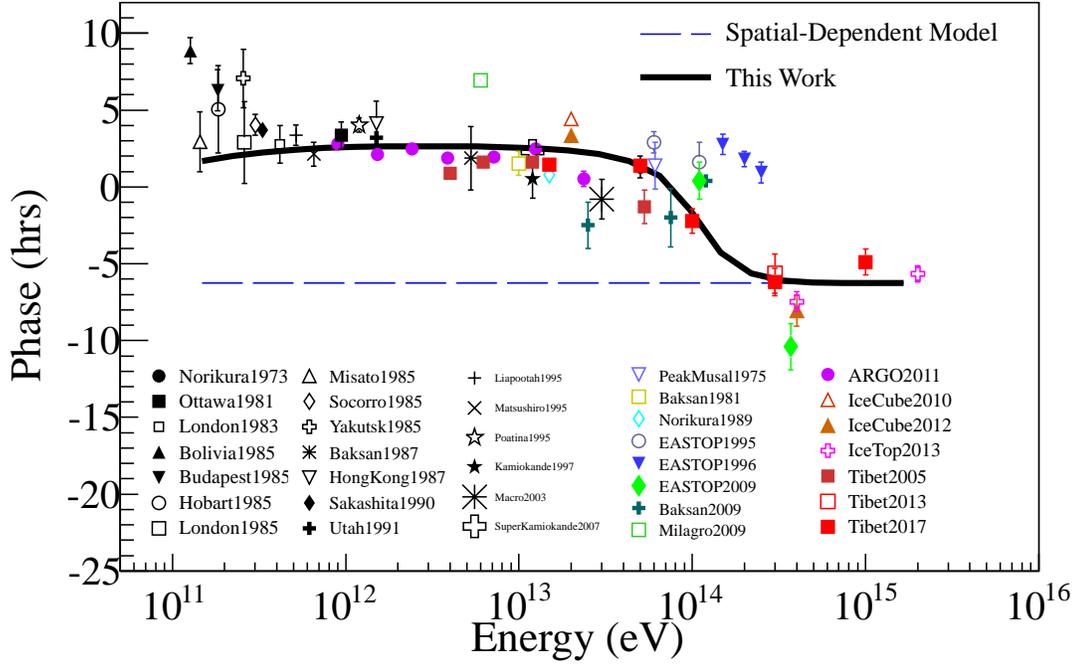}
\caption{\label{fig2}The phase of the DA expected from this work compared
with the observations. The solid black line and blue dash line represent the model calculations in this work and the SDP model, respectively. The data points for the DA phase is token from the same observations as described in Fig.~\ref{fig1}.}
\end{figure}

Figure~\ref{fig1} shows the amplitude of anisotropy comparison between model calculations and the experimental results. The solid black line and blue dash line represent the model calculations in this work and the SDP model, respectively. In the scenario of steady-state propagation, such as the SDP model, the anisotropy grows with diffusion coefficient, namely rising with the energy. In this work a nearby single source is added to the SDP model, the cosmic ray accelerated from this single source will dominate the CR gradient at the energy region below 0.1 PeV, so the anisotropy in this energy will mainly decided by this single source. The dipole anisotropy amplitude is initially increasing with energy below 10 TeV, and begins to decrease above 10 TeV, this behavior is consistent with the single source spectrum. Because the high energy cut-off, while the energy is above 100 TeV the influence of the single source vanished, and the amplitude begins to increase again under the influence of background sources. The phase will also has an abrupt change at the 100 TeV energy scale where the dominator of the anisotropy is changing from the single source to the background sources. As showed in figure~\ref{fig2} the solid black line, the dipole phase change can reproduced by adding a single source to the SDP model. The blue dash line in figure~\ref{fig2} shows the phase of the DA calculated by the SDP model only with the background sources, which is expected to simply align with the direction of the Galactic center.

\subsection{The Spectrum of Proton}

\begin{figure}[ht!]
\plotone{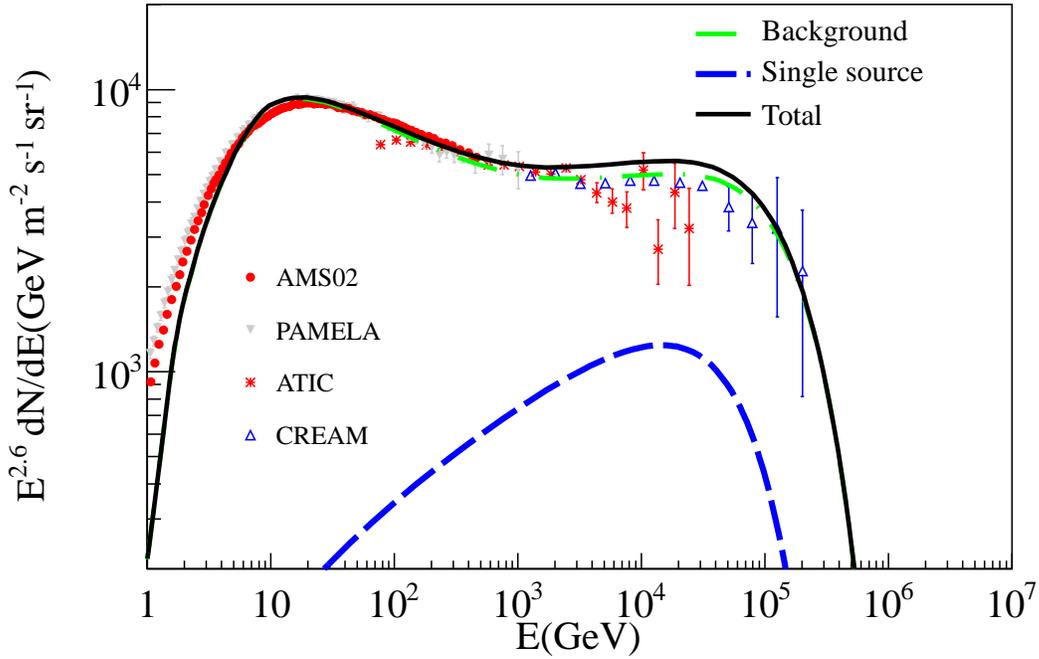}
\caption{\label{fig3}  Comparison between model calculations and observations for the primary spectrum of protons. The experiment data come from: AMS02\citep{gie15}, PAMELA\citep{adr11}, ATIC\citep{pan06}, CREAM\citep{ahn10}.}
\end{figure}

Figure~\ref{fig3} shows the spectrum comparison between model calculations and the experimental results for protons. The green dash-dot, blue dashed and solid black lines represent the calculations for the background sources, nearby single source and total respectively.
It has been discussed in \citet{guo16} that the SDP model can gives clear hardening of the spectrum for E $\geq$ 300 GeV, which is consistent with the data. With the single source added, the proton spectrum can maintain this structure.

\section{Conclusion}

In this study we have shown that the DA amplitude and phase evolution with energies can be reproduce simultaneously when a nearby single source is added to the SDP model. In recent years, the CR anisotropy has been well measured, but the traditional diffusion theory can't explain these new observations, especially the phase flip phenomenon. In this work, the CR DA can be considered to be made of two component: one is induce by nearby single source and the other is caused by all other galactic sources. At the energy region below $10^{14}$ eV, the nearby source can have dominating effect to the CR gradient and decide the phase of the DA, but beyond that, due to the high-energy cutoff of local proton flux, the phase of the DA comes back to the case of SDP model. In other words the DA phase will take a phase flip at the energy of $10^{14}$ eV. With the influence of the single source, the amplitude of the DA is not a simple power law, its evolution with energies can be well reproduced by adding a nearby source to the SDP model.

\acknowledgments

We thank YiQing Guo, Wei Liu, HongBo Hu, SiMing Liu, Qiang Yuan and Yi Zhang for
helpful discussions. This work is supported by the Ministry of Science and Technology of
  China, Natural Sciences Foundation of China (11505291).

\emph{Note added in proof}. While this paper was in preparation, a similar
paper discussing nearby sources effects to the anisotropy and the spectrum under the SDP model scenario appearing on arXiv\citep{liu18}.

\end{document}